\documentclass[journal]{IEEEtran}
\usepackage{color}
\usepackage{array}
\usepackage{booktabs}
\usepackage{cite}
\usepackage{threeparttable}
\usepackage{multirow}

\usepackage{graphicx}
\usepackage{hyperref}
\hypersetup{hidelinks,
	colorlinks=true,
	allcolors=black,
	pdfstartview=Fit,
	breaklinks=true
}
\usepackage{amsmath,bm}
\usepackage{amssymb}
\usepackage{amsthm}
\usepackage[capitalize, nameinlink]{cleveref}
\newtheoremstyle{IEEEtheorem}
  {}
  {}
  {}
  {10pt}
  {\itshape}
  {:}
  { }
  {\thmname{#1}\thmnumber{ #2}\thmnote{ (#3)}}

\theoremstyle{IEEEtheorem}

\newtheorem{thm}{Theorem}
\newtheorem{ass}{Assumption}
\newtheorem{cor}{Corollary}

\newtheorem{rmk}{Remark}
\newtheorem{dfn}{Definition}
\newtheorem{pro}{Proposition}

\usepackage[ruled,linesnumbered]{algorithm2e}

\usepackage[normalem]{ulem}
\renewcommand{\baselinestretch}{1.0} 

\begin{document}
%
\title{Cyber-Resilient Self-Triggered Distributed Control of Networked Microgrids Against Multi-Layer DoS Attacks}
%
%
%

\author{Pudong~Ge,~\IEEEmembership{Student Member,~IEEE,}
       Boli~Chen,~\IEEEmembership{Member,~IEEE} and~Fei~Teng,~\IEEEmembership{Senior Member,~IEEE}

\thanks {This work was supported by EPSRC under Grant EP/W028662/1 and by The Royal Society under Grant RGS/R1/211256.
(Corresponding author:  Dr  Fei  Teng {\tt\small(f.teng@imperial.ac.uk)}).}

\thanks {Pudong Ge and Fei Teng are with the Department of Electrical and Electronic Engineering, Imperial College London, London SW7 2AZ, U.K. {\tt\small(pudong.ge19@imperial.ac.uk; f.teng@imperial.ac.uk)}}
\thanks {Boli Chen is with the Department of Electronic and Electrical Engineering, University College London, London WC1E 6BT, U.K. {\tt\small(boli.chen@ucl.ac.uk)}}
}

\maketitle

\begin{abstract}
Networked microgrids with high penetration of distributed generators have ubiquitous remote information exchange, which may be exposed to various cyber security threats. This paper, for the first time, addresses a consensus problem in terms of frequency synchronisation in networked microgrids subject to multi-layer denial of service (DoS) attacks, which could simultaneously affect communication, measurement and control actuation channels. A unified notion of Persistency-of-Data-Flow (PoDF) is proposed to characterise the data unavailability in different information network links, and further quantifies the multi-layer DoS effects on the hierarchical system. With PoDF, we provide a sufficient condition of the DoS attacks under which the consensus can be preserved with the proposed edge-based self-triggered distributed control framework. In addition, to mitigate the conservativeness of offline design against the worst-case attack across all agents, an online self-adaptive scheme of the control parameters is developed to fully utilise the latest available information of all data transmission channels. Finally, the effectiveness of the proposed cyber-resilient self-triggered distributed control is verified by representative case studies.
\end{abstract}


%
\IEEEpeerreviewmaketitle

\section{Introduction}
\IEEEPARstart{T}{he} energy source has been transforming from traditional fossil fuel based power generations to inverter-based renewable energy resources driven by the development of low/zero-carbon societies~\cite{creutzig2014catching}. Rapidly developing {inverter-based distributed energy resources (DERs)} gradually dominate power systems~\cite{wang2022electrifying,wang2020sustainable}. Reconstructing high-DER-penetrated power systems into multi-microgrids, i.e. networked microgrids (MGs) is one of the significant pathways of improving the resilience~\cite{gholami2021stability,ge2022resilience}. However, the integration of increasing DERs (using the concept of networked MGs) has lead to more complicated information flows and  tighter cyber-physical fusion \cite{pasqualetti_control-theoretic_2015} between DER devices and information systems in order to support efficient control logic. The large scale integration of distributed DERs restricts the applicability of traditional centralised control methods due to the communication constraints and vulnerability against single-point failure, which drives the rapid development of distributed control methods \cite{ge_eventtriggered_2020,ge_resilient_2021}. 

Such cyber-physical system has inevitably left multi-MG systems exposed to uncertainties from the physical environment and malicious cyber attacks from cyberspace. One of the most significant cyber-layer issues is known as denial-of-service (DoS) or jamming attacks, which intend to disrupt communication and data exchange among networked MG information systems to deteriorate control and operation performance.
Therefore, resilient distributed control has been receiving significant attention in recent years. Various control methods have been proposed to enhance the resilience of cyber-physical MGs against DoS attacks, including time-varying sampling strategies \cite{deng2021distributed,lian2021distributed,danzi2019software}, Lyapunov-based analysis \cite{liu2019stochastic,hu2022resilient,wan2021distributed}, $H_\infty$ control \cite{zhang2021attack,hu2020resilient}, switched system design \cite{hu2020attack,liu2021resilient,chlela2018fallback} and reinforcement learning \cite{chen2022multi}. To efficiently manage the information flow, the concept of event-/self-triggered control strategies \cite{heemels_introduction_2012} is developed to enable aperiodic communication, sensing and actuation \cite{dimarogonas_distributed_2012}. With the event-/self-triggered framework, a class of effective DoS countermeasures are designed by constructing suitable triggering mechanisms inferred from Lyapunov arguments \cite{feng_secure_2019,xu2019distributed,danzi2019software,lian2021distributed,senejohnny_jamming-resilient_2018,de_persis_robust_2013}. For instance, the works presented in \cite{danzi2019software,lian2021distributed} propose an adaptive sampling mechanism whereby the impact of DoS attacks can be mitigated by increasing the sampling rate under attacks.

{Existing literature on DoS attacks can be generalised into two categories: 1) attacks only over neighbouring communication links, 2) attacks over the sensing-communication-actuation chain. The neighbouring communication links admittedly are the most vulnerable to attackers as discussed in~\cite{deng2021distributed,lian2021distributed,danzi2019software,wan2021distributed,liu2021resilient,xu2019distributed,senejohnny_jamming-resilient_2018}. Ref. \cite{feng_secure_2019}, though mentioning multi-layer DoS attacks, still focuses on the effects on communication channels. However, the sensing and actuation channels are also worthy of consideration. Some recent works start to investigate the attacks over sensing-communication-actuation chains, by either focusing on the single-layer sensing and actuation channels while ignoring communication channels~\cite{liu2019stochastic}, or simply regarding DoS attack effects on the chains as overdue input updates~\cite{hu2020attack,hu2022resilient,hu2020resilient}. In this context, there is still a lack of understanding of the diverse impact of DoS attacks against different layers of the sensing-communication-actuation chain in a hierarchical control framework of power systems.
}

In fact, a hierarchical control framework adopted by networked MGs relies on more complex information network. On this occasion, each DG involves remote (e.g., telemetered) sensing and control actuation with its MG centre controller (MGCC). Hence, cyber attacks could simultaneously occur on communication links for inter-MG data sharing, measurement and actuation channels for intra-MG aggregation and distribution respectively. In particular, the adversary can erase the data sent to actuators or to block the sensor measurement. This motivates the resilience enhancement against multi-layer DoS for networked MGs within a hierarchical control framework. In this context, this paper proposes a novel scheme that, for the first time, addresses multi-layer DoS attacks targeting the neighbouring communication, sensor measurement and control actuation channels of networked MGs with hierarchically controlled DERs.
The main contributions are summarized as follows:
\begin{enumerate}
	\item To characterise multi-layer DoS attacks within different data flow channels among networked MGs, we propose a unified notion of Persistency-of-Data-Flow (PoDF). The notion PoDF is of significance in evaluating the effects of multi-layer DoS attacks.
	\item With an edge-based control logic, the proposed self-triggered ternary controller enables asynchronous data collection and processing for each MG from all its neighbours as opposed to existing methods in that relays on synchronous communication. This remarkable feature of asynchronous data collection and processing turns out to be of major significance to ensure consensus properties in the presence of multi-layer DoS attacks.
	\item An adaptive scheme of the control and communication policies is devised by utilising timestamps of successful information exchange attempts in different information network links. As such, the conservativeness of the edge-based self-triggered control designed from a global perspective can be significantly reduced.
\end{enumerate}

The remainder of this paper is organized as follows. In \cref{sec:preliminaries}, the cyber-physical model of networked MGs and the self-triggered consensus concept are provided. \cref{sec:3} introduces the adaptive distributed self-triggered consensus controller with reduced conservativeness that is proved to be resilient against multi-layer DoS attacks. Simulation results are presented in \cref{sec:5} and \cref{sec:6} concludes this paper.

\section{Preliminaries and Problem Formulation}
\label{sec:preliminaries}

\begin{figure*}[!htp]
	\centering
	\includegraphics[width=\textwidth]{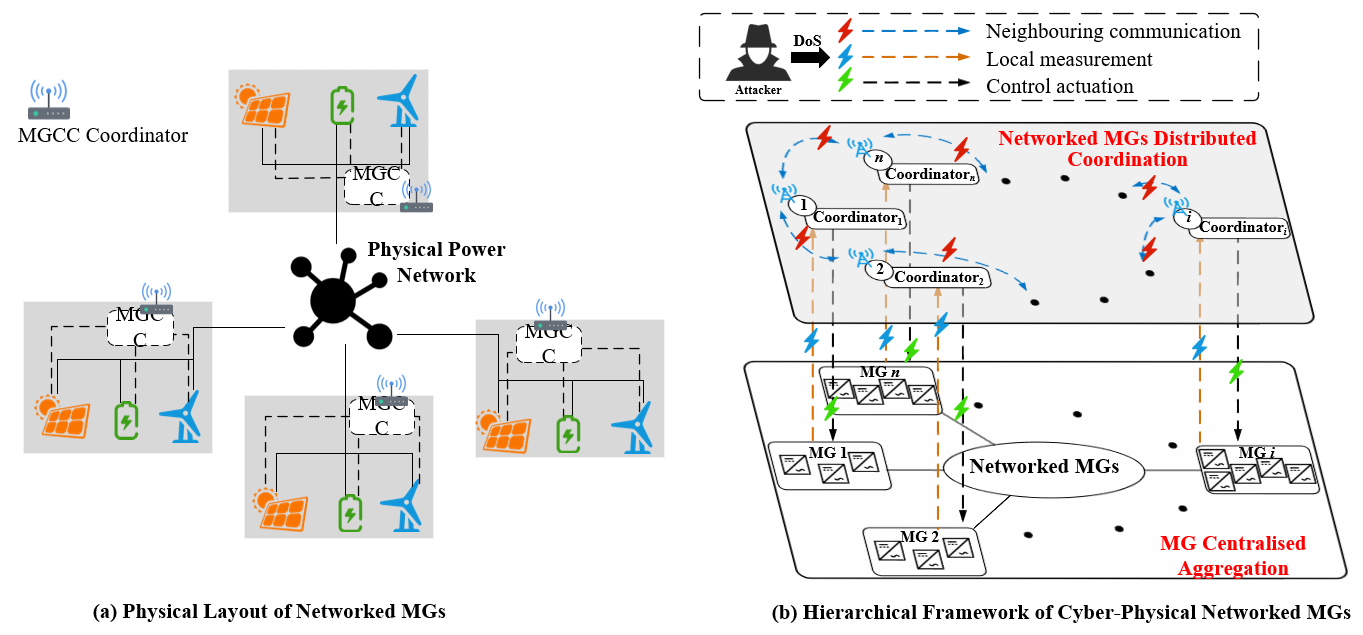}\\[-2ex]
	\caption{{Hierarchically controlled networked MGs.}}
	\label{fig:MGs_diagram}
\end{figure*}

\subsection{Problem Statement}\label{sec:2.1}

The networked MGs discussed in this paper are controlled under a hierarchical framework, as shown in \cref{fig:MGs_diagram}, where each MG employs one central coordinator called MGCC to aggregate the measured information and to distribute the calculated commands. 
{In each MG shown in \cref{fig:MGs_diagram}(a), one MGCC manages all dispatchable DERs and aggregates their operational states. Each MGCC exchanges its own aggregated state information with other neighbouring MGCCs through a distributed communication network, to enable distributed coordination, as depicted in \cref{fig:MGs_diagram}(b).}

The basic idea of such a hierarchical framework is to aggregate DGs, with small capacities but in large quantities inside one MG to support system operation. Such a hierarchical framework \cite{backhaus2016networked} avoids a curse of dimensionality within a fully centralised control, while modularized distributed control avoids the large-scale complex communication network of a fully distributed framework.

To effectively regulate each MG, an aggregated dynamic model can be built through some equivalent methods \cite{chen2021aggregated,roos2020aggregation,gholami2021stability}, even if there exist nodes without DGs (refer to \cite{dorfler2013kron}). To summarise, consider a droop-controlled equivalent modelling, for each MG $i$, we have the equivalent parameters
\begin{align}
	m_{Pi}=\frac{1}{\sum_{j\in\mathcal{C}_{i}}\frac{1}{m_{i}^{Pj}}},\,
	\omega_{i}=\frac{\sum_{j\in\mathcal{C}_{i}}\frac{\omega_{i}^{j}}{\omega_{c}m_{i}^{Pj}}}{\sum_{j\in\mathcal{C}_{i}}\frac{1}{\omega_{c}m_{i}^{Pj}}} \label{eq:agg_model}
\end{align}
where $\mathcal{C}_{i}$ contains all DGs of MG $i$. In MG $i$, $m_{i}^{Pj},\omega_{i}^{j}$ denote the frequency droop coefficient and angular frequency of DG $j$, and $m_{Pi}, \omega_{i}$ are respectively the equivalent frequency droop coefficient and the equivalent angular frequency of MG $i$ (similar to the concept of the Center of Inertia). {$\omega_{c}$ denotes the cut-off frequency of low-pass filter in the inverter control loop.}


The objective is to enable each MG to participate frequency synchronisation using
\begin{align}
	\omega_{ni} = \omega_{i} +m_{Pi}P_{i}
	\label{eq:mg_droop}
\end{align}
where $\omega_{ni}$ is the nominal set point for frequency regulation; $P_{i}$ is the summation of active power output of the $i$th MG.

The primary control through \eqref{eq:mg_droop} can not eliminate the frequency deviations from the reference, and the secondary control is employed to achieve frequency synchronisation and accurate active power sharing, i.e.,
\begin{align}
	\lim_{t\rightarrow\infty}|\omega_{i}-\omega_{j}|=0, \lim_{t\rightarrow\infty}\omega_{i} = \omega_{\rm ref}\label{eq:obj_f}\\
	\lim_{t\rightarrow\infty}\left|\frac{P_{i}}{P_{\max,i}}-\frac{P_{j}}{P_{\max,j}}\right|=0\label{eq:obj_p}
\end{align}
where $P_{\max,i}$ denotes the active power ratings of the $i$th generator, and \eqref{eq:obj_p} is equivalent to $\lim_{t\rightarrow\infty}|m_{Pi}P_{i}-m_{Pj}P_{j}|=0$ by approximately setting frequency droop coefficients.

To formulate the control problem, we differentiate \eqref{eq:mg_droop} and choose the changing rates of frequency and active power output as control variables $\dot{\omega}_{ni}=\dot{\omega}_{i}+m_{Pi}\dot{P}_{i} = u_{\omega i}+u_{Pi}$ {with $u_{\omega i},u_{Pi}$ being the auxiliary control inputs that have been widely utilised in \cite{bidram_multiobjective_2014,dehkordi2017distributed}. Such that, we can obtain}
\begin{align}
	\dot{\bm{x}}_{\omega}=\bm{u}_{\omega},\dot{\bm{x}}_{P}=\bm{u}_{P}\label{eq:f_matrix}
\end{align}
where $\bm{x}_{\omega} = [\omega_{1},\dots,\omega_{n}]^\top$, $\bm{x}_{P} = [m_{P1}P_{1},\dots,m_{Pn}P_{n}]^\top$, $\bm{u}_{\omega} = [u_{\omega 1},\dots,u_{\omega n}]^\top$ and $\bm{u}_{P} = [u_{P1},\dots,u_{Pn}]^\top$. 
Owing to the similar formulation of modelling \eqref{eq:f_matrix} for frequency and active power, we hereafter omit the subscript $\omega,P$, i.e., $x_i:=\omega_i$ or $x_i:=m_{Pi}P_{i}$, to design the control algorithm that can be applied to both frequency regulation and active power sharing.

The communication topology of networked MGs can be modelled by an undirected graph $\mathcal{G=\{I,E\}}$, where $\mathcal{I}=\{1,2,\dots,m\}$ is a set of MGs, $\mathcal{E \subseteq I \times I}$ is a set of edges, and $m$ is the number of MGs. An edge $(j,i)$ means that the $i$th MG can receive information from the $j$th MG and $j$ is a neighbour of $i$. The set of neighbours of MG $i$ is described by $\mathcal{N}_{i}=\{j:(j,i)\in \mathcal{E}\}$ with $d_{i} = |\mathcal{N}_{i}|$ denoting the cardinality of $\mathcal{N}_{i}$. {The corresponding adjacency matrix $\mathcal{A}=[a_{ij}]\in \mathbb{R}^{m \times m}$ is formed by $a_{ii}=0$; $a_{ij}>0$ if $(j,i)\in \mathcal{E}$, otherwise $a_{ij}=0$. The communication topology is denoted by the matrix $\mathcal{A}$, which is assumed to be connected to guarantee the consensus performance~\cite{ren2005survey}.}

As shown in \cref{fig:MGs_diagram}, different channels, i.e., measurement, communication and actuation are vulnerable to cyberattacks due to the hierarchical structure. In this paper, we consider data unavailability issues affecting all channels. Under multi-layer DoS attacks, the frequency synchronisation problem based on dynamics \eqref{eq:f_matrix} becomes: \textit{how to design efficient control laws to update input vectors $\bm{u}_{\omega},\bm{u}_{P}$ to reach both \eqref{eq:obj_f} and \eqref{eq:obj_p} under DoS attacks?}

\subsection{Preliminary of Distributed Ternary Control}
System \eqref{eq:f_matrix} can be recast in the form of \eqref{eq:system dynamic}, which has been addressed in the literature by a distributed ternary control mechanism. Some basic concepts concerning the ternary control are presented below with more detailed discussion in \cite{senejohnny_jamming-resilient_2018} and \cite{de_persis_robust_2013}. The system is formed by a triplet of $n$-dimensional variables $(x,u,\theta)\in\mathbb{R}^{n}\times\mathbb{R}^{d}\times\mathbb{R}^{d}$, where $x,u,\theta$ are the vectors of node states, controls and clock variables respectively. $u,\theta$ are both edge-based variables with $d:=\sum_{i=1}^{n}{d_{i}}$ defined in \cref{sec:2.1}. {The system dynamics of distributed ternary control are governed by:
\begin{align}
	&\begin{aligned}
		\dot{x}_{i}  = u_{i} = \sum_{j\in \mathcal{N}_{i}}u_{ij}
	\end{aligned}\label{eq:system dynamic}\\
	&\left\{
	\begin{aligned}
		& x_i(t)=x_i(t^-)\quad \forall i\in\mathcal{I} \\
		& u_{ij}(t)=\left\{
		\begin{aligned}
			&\mathrm{sign}_\varepsilon\left(D_{ij}(t)\right), &&\mbox{if }(i,j)\in \mathcal{J}(\theta,t)\\
			&u_{ij}(t^-), &&\mbox{otherwise}
		\end{aligned}
		\right.\\
		& \theta_{ij}(t)=\left\{
		\begin{aligned}
			&f_{ij}\left(x(t)\right), &&\mbox{if }(i,j)\in \mathcal{J}(\theta,t)\\
			&\theta_{ij}(t^-), &&\mbox{otherwise}
		\end{aligned}
		\right.
	\end{aligned}
	\right.\label{eq:controller}
\end{align}
where $i\in\mathcal{I}$, $j\in\mathcal{N}_{i}$. The control input $u_{i}$ aggregates contributions of all edges $(j,i)\in\mathcal{E}$, and $u_{ij}$ represents the control action on node $i$ of the communication link from node $j$ to node $i$. Through \eqref{eq:controller}, $u_{ij}, \theta_{ij}$ are updated only when the clock variable $\theta_{ij}$ reaches zero, i.e., 
$(i,j)\in\mathcal{J}(\theta,t)=\{(i,j): j\in\mathcal{N}_{i}\land \theta_{ij}(t^{-})=0\}$
where $\theta_{ij}(t^{-})=\lim_{\tau\rightarrow t}\theta_{ij}(\tau)$.} Specifically,
\begin{align}\begin{aligned}
	&f_{ij}\left(x(t)\right)=\max\left\{{\frac{|D_{ij}(t)|}{2(d_i+d_j)},\frac{\varepsilon}{2(d_i+d_j)}}\right\}\\
	&D_{ij}(t)=x_j(t)-x_i(t)\\
	&\mathrm{sign}_{\varepsilon}(z):=\left\{
		\begin{aligned}
			&\mathrm{sign}(z),&&\mbox{if }|z|\geq\varepsilon\\
			&0, &&\mbox{otherwise}
		\end{aligned}\right.
	\end{aligned}\label{eq:xxxxx}
\end{align}
with $\varepsilon>0$, a user designed sensitivity parameter (consensus error bound); $u_{ij} \in \{-1,0,1\}$ from a quantiser $\mathrm{sign}_{\varepsilon}(z)$.

\section{Resilient Frequency Regulation of MGs Against Multi-Layer DoS Attacks}\label{sec:3}
In this section, we design a DoS-resilient control strategy for global consensus to mitigate the joint impacts of multi-layer DoS attacks in the networked MGs frequency control. We firstly model the multi-layer DoS attacks and analyse the effects on the data flow serving for the frequency regulation. Inspired by the concept of self-triggered control, the adaptive distributed self-triggered control is proposed, and its consensus stability and convergence time are theoretically analysed. Before proposing the DoS-resilient control, we give a comprehensive modelling of multi-layer DoS attacks.

\subsection{Denial-of-Service Attacks Modelling}
To model DoS attacks, $\Xi(t_1,t_2)$ and $\Theta(t_1,t_2)$ are respectively defined as the under-attack and healthy subsets of the time interval $[t_1,t_2)$. By $n(t_1,t_2)$ denoting the incidence of DoS inactive/active transitions within the time interval $[t_1,t_2)$, the following assumption are introduced~\cite{senejohnny_jamming-resilient_2018,feng_secure_2019}, where a more comprehensive information on DoS frequency and duration is provided.
\begin{ass}[DoS Frequency and Duration]
	There exist $\eta \in \mathbb{R}_{\geq 0},\kappa \in \mathbb{R}_{\geq 0}$ and $\tau^{f} \in \mathbb{R}_{\geq 0},\tau^{d} \in \mathbb{R}_{\geq 0}$ such that
	\begin{align}
		&\mathrm{Frequency:}\quad n(t_1,t_2)\leq \eta + \frac{t_2-t_1}{\tau^f}, \label{eqn:DoS frequency}\\
		&\mathrm{Duration:}\quad |\Xi(t_1,t_2)|\leq \kappa + \frac{t_2-t_1}{\tau^d}. \label{eqn:DoS duration}
	\end{align}
	\label{assum:DoS}
\end{ass}\vspace{-1em}

To model multi-layer DoS attacks in a unified form, the Persistency-of-Communication (PoC) in \cite{senejohnny_jamming-resilient_2018} is generalized and extended to a notion of PoDF owing to the independence of DoS on diverse channels of data transmission.

\begin{pro}[Persistency-of-Data-Flow (PoDF)]
	For any transmission channel $\mu\in\{\mathcal{I}\cup\mathcal{E}\}$\footnote{$\mu:=ij$, communication channel $(i,j)\in\mathcal{E}: j\in\mathcal{N}_{i}$; $\mu:=i$, measurement channel of subsystem $i\in\mathcal{I}$; $\mu:=i0$, control actuation channel of subsystem $i\in\mathcal{I}$.} serving for the distributed control, if multi-layer DoS sequences satisfy \cref{assum:DoS} respectively with coefficients $\tau_{\mu}^f$, $\tau_{\mu}^d$, such that
	$
		\phi_{\mu}(\tau_{\mu}^{f},\tau_{\mu}^{d},\Delta_{\mu}^{*}):=\frac{1}{\tau_{\mu}^{d}}+\frac{\Delta_{\mu}^{*}}{\tau_{\mu}^{f}}<1
	$,
	where $\Delta_{\mu}^{*}:=\min\Delta_{\mu}$. Then, for any unsuccessful data transmission attempt $t_{\mu}^k$, at least one successful transmission occurs within the time interval $[t_{\mu}^k,t_{\mu}^k+\Phi_{\mu}]$ with
	$
		\Phi_{\mu}:=\frac{\kappa_{\mu}+(\eta_{\mu}+1)\Delta_{\mu}^{*}}{1-\phi_{\mu}(\tau_{\mu}^{f},\tau_{\mu}^{d},\Delta_{\mu}^{*})}\,.
	$
	\label{prop:PoDF}
\end{pro}
\begin{IEEEproof}
	The proof is similar to that in the Appendix A of~\cite{senejohnny_jamming-resilient_2018}, thus omitted here.
\end{IEEEproof}
\cref{prop:PoDF} describes the impact of multi-layer DoS attacks on each data flow channel. $\Delta_{\mu}^{*}$ denotes the minimum time interval between two sequential attempts of data flow, which is different for the three different types of data transmissions. In practice, {$\Delta_{\mu}^{*}$ can be known a priori}, though conservatively, based on the specification of the system. More specifically, $\Delta_{i}^{*},\Delta_{i0}^{*}$ depend on the performance of each MGCC, while $\Delta_{ij}^{*}$ is determined by \eqref{eq:minimum communication attempt}, which is introduced later.

\begin{ass}\label{assum:multi-layer_DoS_td_tf}
	Assuming that both local-level DoS attacks (measurement and control actuation DoS) occur with similar chance, which is less frequent than that on the neighbouring communication channels, {such that $\tau_{i}^{f}\approx\tau_{i0}^{f},\tau_{i}^{d}\approx\tau_{i0}^{d}\Longrightarrow\Phi_{i}\approx\Phi_{i0}$ and $\Phi_{i}\leq\Phi_{ij},\Phi_{i0}\leq\Phi_{ij}$ according to the definition in \cref{prop:PoDF} and its footnotes.}
\end{ass}

\subsection{DoS Resilient Consensus Control Algorithm}

The distributed control protocol \eqref{eq:system dynamic}--\eqref{eq:xxxxx} is based on the hypothesis that the MGCC has access to both local state $x_{i}(t)$ and neighbouring state $x_{j}(t)$ at the triggering time, and therefore not valid for multi-layer DoS attacks. To ensure the cyber-resilient consensus in such a scenario, we design an adaptive self-triggered control protocol to achieve resilience under multi-layer DoS attacks (the corresponding stability criteria will be discussed later in \cref{section: global stability} and \cref{sec:conservative}). The nominal discrete transition \eqref{eq:controller} is modified as follows:
\begin{align}
	\left\{
	\begin{aligned}
		& x_i(t)=x_i(t^-)\quad \forall i\in\mathcal{I} \\
		& u_{ij}(t)\!=\!\left\{
		\begin{aligned}
			&\mathrm{sign}_\varepsilon\left(D_{ij}(\bar{t})\right), &&(i,j)\!\in\!\mathcal{J}(\theta,t)\!\land\! t \!\in\!\Theta_{ij}(0,t)\\
			&0, &&(i,j)\!\in\!\mathcal{J}(\theta,t)\!\land\! t\!\in\!\Xi_{ij}(0,t)\\
			&u_{ij}(t^-), && \mbox{otherwise}
		\end{aligned}
		\right.\\
		& \theta_{ij}(t)\!=\!\left\{
		\begin{aligned}
			&f_{ij}\left(x(\bar{t})\right), &&(i,j)\!\in\!\mathcal{J}(\theta,t)\!\land\! t \!\in\!\Theta_{ij}(0,t)\\
			&\frac{ \varepsilon_{ij}}{2(d_i+d_j)}, &&(i,j)\!\in\!\mathcal{J}(\theta,t)\!\land\! t\!\in\!\Xi_{ij}(0,t)\\\\
			&\theta_{ij}(t^-), && \mbox{otherwise}
		\end{aligned}
		\right.
	\end{aligned}
	\right.\label{eq:adaptive controller}
\end{align}
with asynchronous clock rate across all network links $\dot{\theta}_{ij}(t) = -R_{ij}$ and individual sensitivity parameters $\varepsilon_{ij}$ satisfying: 
\begin{align}
    0<\varepsilon\leq\varepsilon_{ij}.
    \label{eq:low bound varepsilon}
\end{align}
where $\varepsilon$ represents the minimally acceptable consensus error that avoids Zeno-behaviour of all edges.
The utilization of $R_{ij}$ and $\varepsilon_{ij}$, for each edge as opposed to the uniform parameters used in the nominal scheme \eqref{eq:system dynamic}-\eqref{eq:xxxxx} is a remarkable feature, and it turns out to be useful in the context of consensus performance as will be discussed in \cref{sec:conservative}. The map $f_{ij}:\mathbb{R}^2\rightarrow\mathbb{R}_{>0}$ is defined as 
$
	f_{ij}\left(x(\bar{t})\right)=\max\left\{{\frac{|D_{ij}(\bar{t})|}{2(d_i+d_j)},\frac{\varepsilon_{ij}}{2(d_i+d_j)}}\right\}\,.
$

Let $\{t_{ij}^{k}\}_{k\in\mathbb{Z}_{\geq0}}$ be the sequence of communication-triggering attempt. It is immediate to show that a dwell-time property is ensured between consecutive sequences:
\begin{align}
	\Delta_{ij}:=t_{ij}^{k+1}-t_{ij}^{k}\geq\frac{\varepsilon_{ij}}{2R_{ij}(d_i+d_j)}\geq\frac{\varepsilon}{4R_{ij}d_{\max}}
	\label{eq:minimum communication attempt}
\end{align}
where $d_{\max}=\max_{i\in\mathcal{I}}d_{i}$. This ensures the adaptive self-triggered control \eqref{eq:adaptive controller} to be Zeno-free. 
The item $D_{ij}(\bar{t})$ of \eqref{eq:adaptive controller} is designed to mitigate the cooperative impacts of multi-layer DoS, i.e.,
$D_{ij}(\bar{t})=x_j(\bar{t}_j)-x_i(\bar{t}_i)$,
where ``$\bar{t}$" denotes latest time instant when the state is available. 

For the sake of further analysis, we define
\begin{dfn}[Secure Consensus]\label{defn:secure consensus}
	Given the system \eqref{eq:system dynamic}, a graph $\mathcal{G}$ and a distributed self-triggered resilient consensus controller with edge-based control $u_{ij}$, the networked systems are said to be consensus under multi-layer DoS attacks if for any initial condition, $x(t)$ converges in finite time to a point belonging to the set by defining $\delta=\varepsilon(n-1)$
	\begin{align}
		\{x\in\mathbb{R}^{n}: |x_i(t)-x_j(t)|<\delta\quad\forall (i,j)\in\mathcal{I}\times\mathcal{I}\}.
		\label{eq:convergence set}
	\end{align}
\end{dfn}
\begin{rmk}
	The consensus error bound of the distributed system $\delta$ derives from edges and can be designed appropriately as small as possible to ensure the system consensus performance, i.e., frequency regulation and active power sharing accuracy, just for being Zeno-free.
\end{rmk}
In the following, the distributed control system stability will be analysed in terms of parameter design, followed by the convergence analysis in line with \eqref{eq:convergence set}. The network behaviour of the networked system \eqref{eq:system dynamic}, \eqref{eq:adaptive controller}-\eqref{eq:minimum communication attempt} is analysed in the presence of multi-layer DoS attacks. The analysis is carried out in two steps: 1) we assume uniform clock rate and consensus error bound, such that $R_{ij}=R$, $\varepsilon_{ij}=\varepsilon,\,\forall i\in\mathcal{I}, \, j \in \mathcal{N}_{i}$ and provide the stability condition in a global sense, and 2)  with the additional degrees of freedom endowed by $\varepsilon_{ij}$ and $R_{ij}$, we provide less conservative design criteria by which the consensus remains guaranteed.

\subsection{Control Parameter Design and Stability Analysis}\label{section: global stability} 
After the MGCC $i$ updates the associated input $u_{ij}$ related to its neighbour $j$ by \eqref{eq:adaptive controller}, its transmission through the actuation channels could also be blocked due to DoS attacks. To better demonstrate the effects of DoS attacks on the actuation channels, two sequences of time instants for any $(i,j)\in \mathcal{E}$ are defined: $\{t_{ij}^{k}: k\in\mathbb{N}\}$ and $\{s_{ij}^{k}: k\in\mathbb{N}\}$. The sequence $t_{ij}^{k}$ denotes the time instants at which both local and neighbouring states are updated after $(i,j)\in \mathcal{J}(\theta,t)$ satisfies, while the sequence $s_{ij}^{k}$ denotes the corresponding time instants at which transmission attempts of control actuation from \eqref{eq:adaptive controller} are successful.
Then, two sequences have the property of $0 \leq s_{ij}^{k}-t_{ij}^{k} \leq \Phi_{i0}$.

\begin{thm}\label{thm:stability criteria}
	Consider the distributed control system \eqref{eq:system dynamic}, \eqref{eq:adaptive controller} subject to multi-layer DoS attacks. If \cref{assum:DoS} and \cref{assum:multi-layer_DoS_td_tf} hold and 
	\begin{align}
		\varepsilon>2d_{\max}\Phi_{\mathcal{I}+2\mathcal{I}0}^{\max},\,
		R>\frac{\varepsilon}{2\left[\varepsilon-2d_{\max}\Phi_{\mathcal{I}+2\mathcal{I}0}^{\max}\right]}
		\label{eq:stability criteria}
	\end{align}
	with $\Phi_{\mathcal{I}+2\mathcal{I}0}^{\max}=\Phi_{\mathcal{I}}^{\max}+2\Phi_{\mathcal{I}0}^{\max}$, $\Phi_{\mathcal{I}}^{\max}=\max_{i\in\mathcal{I}}{\Phi_{i}},\Phi_{\mathcal{I}0}^{\max}=\max_{i\in\mathcal{I}}{\Phi_{i0}}$,
	then $x(t)$ reaches consensus in finite time as described in \eqref{eq:convergence set}.
\end{thm}
\begin{IEEEproof}
	Consider any time $t$, there exists two successive time instants of successful control actuation that satisfy $s_{ij}^{k}\leq t< s_{ij}^{k+1}$.
	During the time period $[s_{ij}^{k},s_{ij}^{k+1})$, the control input that is updated through \eqref{eq:adaptive controller} at the time instant $t_{ij}^{k}$ will be applied. For each $(i,j)\in\mathcal{E}:j
	\in\mathcal{N}_{i}$, we have the following inequality:
	\begin{align}\label{eq:time relaxation for proof}
		t-t_{ij}^{k}\leq \frac{f_{ij}(x(\bar{t}_{ij}^{k}))}{R}+2\Phi_{i0}
	\end{align}
	Then if $D_{ij}(\bar{t}_{ij}^{k})\geq \varepsilon$,
	\begin{align}
		\begin{aligned}
			& D_{ij}(t)=x_j(t)-x_i(t) \\
			& \overset{(a1)}{\geq} D_{ij}(t_{ij}^{k})-(d_i+d_j)(t-t_{ij}^{k})\\
			& \overset{(a2)}{\geq} D_{ij}(\bar{t}_{ij}^{k})-d_i\Phi_{i}-d_j\Phi_{j}-(d_i+d_j)(t-t_{ij}^{k}) \\
			& \overset{(a3)}{\geq} D_{ij}(\bar{t}_{ij}^{k})(1\!-\!\frac{1}{2R})\!-\!d_{i}(\Phi_{i}\!+\!2\Phi_{i0})\!-\!d_{j}(\Phi_{j}\!+\!2\Phi_{i0})
		\end{aligned}\label{eq:local slack}
	\end{align}
	where (a1) derives from identifiable neighbours and control inputs, and (a2), (a3) are from Proposition~\ref{prop:PoDF} and \eqref{eq:time relaxation for proof} respectively, then \eqref{eq:local slack} can be expressed as
	\begin{align}
		\begin{aligned}\label{eq:Dij}
			D_{ij}(t)\geq  D_{ij}(\bar{t}_{ij}^{k})(1-\frac{1}{2R})-2d_{\max}\Phi_{\mathcal{I}+2\mathcal{I}0}^{\max}>0
		\end{aligned}
	\end{align}
	If $D_{ij}(\bar{t}_{ij}^{k})\leq-\varepsilon$, an analogous inequality holds
	\begin{align}
		\begin{aligned}\label{eq:Dij_negative}
			D_{ij}(t)\leq  D_{ij}(\bar{t}_{ij}^{k})(1-\frac{1}{2R})+2d_{\max}\Phi_{\mathcal{I}+2\mathcal{I}0}^{\max}<0
		\end{aligned}
	\end{align}
	Define error terms as $e_i=x_i-\frac{1}{n}\sum_{i=1}^{n}x_i$ and $\boldsymbol{e}=[e_i]_{N\times1}$. Consider a candidate Lyapunov function $V(t)=\frac{1}{2}\boldsymbol{e}^T\boldsymbol{e}$ and define $\mathcal{S}:=|D_{ij}(\bar{t}_{ij}^{k})|\geq\varepsilon\land t_{ij}^{k}\in \Theta_{ij}(0,t)$, then the derivative of $V(t)$ under the controller \eqref{eq:adaptive controller}:
	\begin{align}
		\begin{aligned}
			\dot{V}(t)&=\sum_{i=1}^{n}e_{i}\dot{e}_{i}=\sum_{i=1}^{n}{e_{i}\sum_{j\in\mathcal{N}_{i}: \mathcal{S}}\mathrm{sign}_\varepsilon\left(D_{ij}(\bar{t})\right)}\\
			&=-\frac{1}{2}\sum_{(i,j)\in\mathcal{E}:\mathcal{S}}D_{ij}(t)\mathrm{sign}_\varepsilon\left(D_{ij}(\bar{t})\right)\\
			&\leq -\frac{1}{2}\sum_{(i,j)\in\mathcal{E}:\mathcal{S}}\Bigg[\varepsilon\left(1-\frac{1}{2R}\right) \\
			&\hspace{1cm}-2d_{\max}(\Phi_{\mathcal{I}}^{\max}+2\Phi_{\mathcal{I}0}^{\max})\Bigg]\overset{(b)}{<}0
		\end{aligned}\label{eq:stability proof}
	\end{align}
	where (b) derives by applying \eqref{eq:stability criteria} in Theorem~\ref{thm:stability criteria}. As a result, \eqref{eq:stability proof} shows the convergence {of \cref{thm:stability criteria}}. Thus, \textit{secure consensus} defined in \cref{defn:secure consensus} can be reached.
\end{IEEEproof}

{Based on the results stated in Theorem~\ref{thm:stability criteria}, the convergence time can be characterised.

\begin{cor}[Convergence Time]\label{cor: convergence analysis}
		Consider $T_{\star}$ as the convergence time of the distributed control system \eqref{eq:system dynamic}, \eqref{eq:adaptive controller}. It holds that
	\begin{align}
		T_{\star}\leq\frac{2\varepsilon(d_{\max}+d_{\min})+8Rd_{\max}d_{\min}\Phi_{\mathcal{IJ}+2\mathcal{I}0}^{\max}}{\varepsilon{d_{\min}}\left[\varepsilon(1-\frac{1}{2R})-2d_{\max}\Phi_{\mathcal{I}+2\mathcal{I}0}^{\max}\right]}V(0)
	\end{align}
	where $\Phi_{\mathcal{IJ}+2\mathcal{I}0}^{\max} = \Phi_{\mathcal{IJ}}^{\max}+2\Phi_{\mathcal{I}0}^{\max}$, $\Phi_{\mathcal{IJ}}^{\max} = \max_{i\in\mathcal{I},j\in\mathcal{N}_{i}}{\Phi_{ij}}$, $d_{\min} = \min_{i\in\mathcal{I}}d_{i}$.
\end{cor}
\begin{IEEEproof}
	Consider the Lyapunov function based stability analysis \eqref{eq:stability proof}, for any successful communication attempt $t_{ij}^{k}$ with $|D_{ij}(\bar{t}_{ij}^{k})|\geq\varepsilon$, the function $V$ decreases at least with the rate of $\rho:=\frac{1}{2}\left[\varepsilon(1-\frac{1}{2R})-2d_{\max}\Phi_{\mathcal{I}+2\mathcal{I}0}^{\max}\right]$ by at least $(\varepsilon/4Rd_{\max})$ units of time (as inferred from \eqref{eq:minimum communication attempt}) under the enhanced adaptive controller \eqref{eq:adaptive controller}. 
	
	We consider any $t>0$ the consensus has not yet been reached and $u_{ij}^{\star}(t)=0$, thus the next communication attempt through edge $(i,j)\in\mathcal{E}$ will occur at the following time period $\left[t,t+\varepsilon/4Rd_{\min}\right]$. The most conservative scenario is that over this time period $u_{ij}^{\star}=0$. Due to the effect of DoS on communication channels, one successful communication attempt will certainly occurs before $(t+\varepsilon/4Rd_{\min}+\Phi_{ij})$ even at the most conservative scenario.
	
	Then, we consider the effect of DoS on control actuation channels. After $u_{ij}$ is updated at $t_{ij}^{k}$, the successful control actuation attempt $u_{ij}^{\star}(s_{ij}^{k})=u_{ij}(\bar{t}_{ij}^{k})$ occurs at $s_{ij}^{k}\in\left[t_{ij}^{k},t_{ij}^{k}+\Phi_{i0}\right]$. The time-duration of $u_{ij}^{\star}(s_{ij}^{k})$ contributing to the consensus is determined by the next successful control actuation attempt, which can be defined as $s_{ij}^{k+1}\in\left[t_{ij}^{k+1},t_{ij}^{k+1}+\Phi_{i0}\right]$. We assume $u_{ij}^{\star}(s_{ij}^{k})$ will be lasting for at least $(\varepsilon/4Rd_{\max}+\Delta{t})$ with $0\leq\Delta{t}\leq\Phi_{i0}$, thus, we conclude that $V$ decreases by at least $\left[\rho(\varepsilon/4Rd_{\max}+\Delta{t})\right]$ every $\left(\Phi_{ij}+\varepsilon/4Rd_{\min}+\varepsilon/4Rd_{\max}+\Delta{t}\right)$ units of time. Therefore, the convergence time
	\begin{align}
		\begin{aligned}
			T_{\star}&\leq\frac{\varepsilon/4Rd_{\min}+\Phi_{ij}+\Phi_{i0}+\varepsilon/4Rd_{\max}+\Delta{t}}{\rho(\varepsilon/4Rd_{\max}+\Delta{t})}V(0)\\
			&\leq\frac{\varepsilon/4Rd_{\min}+\Phi_{ij}+2\Phi_{i0}+\varepsilon/4Rd_{\max}}{\rho\varepsilon/4Rd_{\max}}V(0)\\
			&\leq\frac{2\left(\varepsilon/4Rd_{\min}+\varepsilon/4Rd_{\max}+\Phi_{\mathcal{IJ}}^{\max}+2\Phi_{\mathcal{I}0}^{\max}\right)}{\left[\varepsilon(1-\frac{1}{2R})-2d_{\max}(\Phi_{\mathcal{I}}^{\max}+2\Phi_{\mathcal{I}0}^{\max})\right]\varepsilon/4Rd_{\max}}V(0)\\
			&=\frac{2\varepsilon(d_{\max}+d_{\min})+8Rd_{\max}d_{\min}(\Phi_{\mathcal{IJ}}^{\max}+2\Phi_{\mathcal{I}0}^{\max})}{\varepsilon{d_{\min}}\left[\varepsilon(1-\frac{1}{2R})-2d_{\max}(\Phi_{\mathcal{I}}^{\max}+2\Phi_{\mathcal{I}0}^{\max})\right]}V(0)
		\end{aligned}
	\end{align}
\end{IEEEproof}
}

\subsection{Conservativeness Mitigation under DoS Attacks}
\label{sec:conservative}
The global consensus criteria \eqref{eq:stability criteria} given in Theorem~\ref{thm:stability criteria}, though can be designed offline, are inferred from the global worst case analysis in terms of PoDF (uniform bounds across all the MGCC nodes), thereby being conservative and could lead to degraded consensus accuracy. In this section, under the procedure of DoS resilient control protocol summarised in Algorithm~\ref{alg:1}, less conservative criteria are derived from a local perspective (Theorem~\ref{thm:local stability criteria}) to further improve the control performance. 

\begin{algorithm}[!ht]
	\label{alg:1}
	\LinesNumbered
	\caption{DoS Resilient Distributed Consensus Control}
	\textbf{Initialisation:} for all $i\in \mathcal{I}$ and $j\in \mathcal{N}_{i}$, set $\theta_{ij}(0^{-})=0$, $u_{ij}(0^{-})=0$, $u^{\star}_{ij}(0^{-})=0$\;
	\tcc{Local State Update from Sensors to Controllers}
	\ForEach{$i\in\mathcal{I}$}{
		\If{$t\in\Theta_{i}(0,t)$}{
			$i$ updates $x_i(\bar{t})=x_i(t)$\;
		}
	}
	\tcc{Edge-Based Control Update in Controllers}
	\ForEach{$i\in \mathcal{I}$}{
		\ForEach{$j\in \mathcal{N}_{i}$}{
			\While{$\theta_{ij}(t)>0$}{
				$i$ applies the control $u_{i}(t)=\sum_{j\in \mathcal{N}_{i}}u_{ij}(t)$\;
			}
			\uIf{$\theta_{ij}(t)\leq0\land t\in \Theta_{ij}(0,t)$}{
				$i$ updates $u_{ij}(t)=\mathrm{sign}_{\varepsilon}(D_{ij}(\bar{t}))$\;
				$i$ updates $\theta_{ij}(t)=f_{ij}(x(\bar{t}))$\;
			}\ElseIf{$\theta_{ij}(t)\leq0 \land t\in \Xi_{ij}(0,t)$}{
				$i$ updates $u_{ij}(t)=0$\;
				$i$ updates $\theta_{ij}(t)=\frac{\varepsilon_{ij}}{2(d_i+d_j)}$\;
			}
		}
	}
	\tcc{Control Actuation}
	\ForEach{$i\in \mathcal{I}$}{
		\If{$u_i(t)$ is updated $\land\ t\in\Theta_{i0}(0,t)$}{
			$u^{\star}_i(t)=u_i(t)$\;
		}
	}
	\tcp{\textbf{note:} $u_i(t)$ denotes the desired control output, while $u^{\star}_i(t)$ denotes the actual control input of the subsystem. $u_i(t)=u^{\star}_i(t)$ if the actuation channel is not attacked.}
\end{algorithm}

%

\begin{thm}\label{thm:local stability criteria}
Consider the distributed system \eqref{eq:system dynamic} subject to multi-layer DoS attacks and the edge-based control \eqref{eq:adaptive controller}. If each subsystem can individually choose its parameters $\varepsilon_{ij}$ and $R_{ij}$, such that $\forall\,i\in\mathcal{I},\forall\,j\in\mathcal{N}_{i}$,
	\begin{align}
		\begin{aligned}
			\varepsilon_{ij}&>d_{i}(\Phi_{i}+2\Phi_{i0})+d_{j}(\Phi_{j}+2\Phi_{i0})\\
			R_{ij}&>\frac{\varepsilon_{ij}}{2\left[\varepsilon_{ij}-d_{i}(\Phi_{i}+2\Phi_{i0})-d_{j}(\Phi_{j}+2\Phi_{i0})\right]}
		\end{aligned}\label{eq:local stability criteria}
	\end{align}
	then the global consensus \eqref{eq:convergence set} can be guaranteed.
\end{thm}
\begin{IEEEproof}
    See Appendix~\ref{app:thm2}.
\end{IEEEproof}

For the reason that the cyber vulnerability of different links may vary, there exists $\Phi_{i}\leq\Phi_{\mathcal{I}}^{\max},\Phi_{i0}\leq\Phi_{\mathcal{I}0}^{\max},\forall i\in\mathcal{I}$, thus the condition \eqref{eq:local stability criteria} is less conservative than \eqref{eq:stability criteria}. 
Furthermore, although Proposition~\ref{prop:PoDF} gives bounded time interval $\Phi_{\mu}$ that can be utilized to design parameters, not every attack attempt leads to the worst data flow block, i.e., the time to achieve a successful data flow would not be $\Phi_{\mu}$ all the time. Using the bounds to stabilise the system as Theorem~\ref{thm:stability criteria} may lead to excessive conservativeness. Therefore, a self-adaptive scheme is utilised to mitigate the conservativeness.

For the controller of each subsystem $i$, assume the $k$th communication attempt is successful at $t_{ij}^{k}$, we define the following time instants:
\begin{align}
		t_{i,i}^{k} := t_{ij}^{k}-\bar{t}_{i}^{k},\quad t_{i,j}^{k} := t_{ij}^{k}-\bar{t}_{j}^{k},\quad t_{i0}^{k} := s_{ij}^{k}-t_{ij}^{k} \label{eq:timestamp}
\end{align}
where $t_{i,i}^{k},t_{i,j}^{k}$ are available at $t_{ij}^{k}$ whereas $t_{i0}^{k}$ is not know until $t=s_{ij}^{k}$. To estimate $t_{i0}^{k}$, let us consider an unsuccessful control actuation attempt at $\breve{s}_{ij}\in[t_{ij}^{k},s_{ij}^{k})$ and $\hat{t}_{i0}^{k}$ the estimate of $t_{i0}^{k}$. As we know that the next attempt will be made at $\breve{s}_{ij}+\Delta_{i0}^{*}$, 
we keep updating $\hat{t}_{i0}^{k}$ via $\hat{t}_{i0}^{k}=\breve{s}_{ij}+\Delta_{i0}^{*}-t_{ij}^{k}$ until the next successful attempt. As such, there always exists a time instant $\bar{t} < s_{ij}^{k}$, such that for all $t\in [\bar{t}, s_{ij}^{k})$, $\hat{t}_{i0}^{k} = t_{i0}^{k}$. It implies that $t_{i0}^{k}$ is known prior to $s_{ij}^{k}$.

\begin{pro}\label{prop:for self-adaptive}
	For any control actuation during $[s_{ij}^{k},s_{ij}^{k+1})$, the following control inputs are equivalent to the system:
	\begin{align}
		\begin{aligned}
			& u'_{ij}(t)=\mathrm{sign}_\varepsilon\left(D_{ij}(\bar{t}_{ij}^{k})\right)\frac{\vartheta_{ij}^{k}}{\vartheta_{ij}^{k}+\Phi_{i0}},s_{ij}^{k}\leq t<s_{ij}^{k+1}\\
			&\Longleftrightarrow u_{ij}(t)=\left\{\begin{aligned}
				&\mathrm{sign}_\varepsilon\left(D_{ij}(\bar{t}_{ij}^{k})\right),&&s_{ij}^{k}\leq t<t_{ij}^{k*}\\
				&0,&&t_{ij}^{k*}\leq t<s_{ij}^{k+1}
			\end{aligned}
			\right.
		\end{aligned}\label{eq:pro2}
	\end{align}
where $\vartheta_{ij}^{k}=\frac{\theta_{ij}^{k}}{R_{ij}^{k}}=\frac{f_{ij}(x(\bar{t}_{ij}^{k}))}{R_{ij}^{k}}$ and $s_{ij}^{k}+\frac{(\vartheta_{ij}^{k})^{2}}{\vartheta_{ij}^{k}+\Phi_{i0}}\leq t_{ij}^{k*}\leq t_{ij}^{k+1}$.
\end{pro}
\begin{IEEEproof}
    See Appendix~\ref{app:prop2}.
\end{IEEEproof}

Although the consensus error bound $\varepsilon_{ij}$ guaranteed in Theorem~\ref{thm:local stability criteria} is less conservative than \eqref{eq:stability criteria}, it still relies on the PoDF conditions, which is inevitably conservative. Next, we show that a tighter consensus error bound can be achieved if an online self-adaptation mechanism of $\varepsilon_{ij}$ and $R_{ij}$ is permitted after each successful communication attempt.

\begin{cor}[Self-Adaptive Scheme]\label{cor:dynamic edge-based self-triggered strategy}
	Consider the distributed system \eqref{eq:system dynamic} subject to multi-layer DoS attacks and the edge-based control \eqref{eq:adaptive controller} with control input $u'_{ij}$ in Proposition~\ref{prop:for self-adaptive}, if $\varepsilon_{ij}$ and $R_{ij}$ can be adapted after each successful communication attempt, such that 
	\begin{align}
		&\varepsilon_{ij}^{k}>\Gamma_{ij}^{k},\,R_{ij}^{k}>\frac{\varepsilon_{ij}^{k}}{2\left[\varepsilon_{ij}^{k}-\Gamma_{ij}^{k}\right]}\label{eq:self adaptive}
	\end{align}
	where $\Gamma_{ij}^{k}=d_{i}(t_{i,i}^{k}+t_{i0}^{k})+d_{j}(t_{i,j}^{k}+t_{i0}^{k})$ with $t_{i,i}^{k},\,t_{i0}^{k},\,t_{i,j}^{k}$ defined in \eqref{eq:timestamp}, then the secure consensus condition \eqref{eq:convergence set} can be preserved. 
\end{cor}
\begin{IEEEproof}
    See Appendix~\ref{app:cor1}.
\end{IEEEproof} 

After the $k$th successful communication attempt of edge $(i,j)\in\mathcal{E}:j\in\mathcal{N}_{i}$, $\Gamma_{ij}^{k}$ is already known before the control actuation attempt. Then we can choose appropriate $\varepsilon_{ij}^{k},R_{ij}^{k}$ to satisfy \eqref{eq:self adaptive}, and the corresponding clock variable $\theta_{ij}^{k}$ and control variable $u_{ij}^{k}=u'_{ij}$ can be obtained from \eqref{eq:adaptive controller} and \eqref{eq:pro2} respectively. 
To make the proposed self-adaptive scheme clear, we summarise it in Algorithm~\ref{alg:2}.

\begin{rmk}
The conditions shown in \eqref{eq:self adaptive} are equivalent to $\varepsilon_{ij}^{k}>\left[1+\frac{1}{2R_{ij}^{k}-1}\right]\Gamma_{ij}^{k},R_{ij}^{k}>0.5$, which explicitly shows the relationship between two designed parameters. The selection of $\varepsilon_{ij}^{k},R_{ij}^{k}$ is subject to a trade-off between consensus accuracy and computation burden. More specifically, smaller $\varepsilon_{ij}^{k}$ leads to more accurate consensus performance  in terms of \eqref{eq:low bound varepsilon} but requires larger $R_{ij}^{k}$, which means more frequent communication between MGCCs. 
Hence, the parameter selection in practice should consider both the communication capability and accuracy requirement of networked MGs case-by-case.
\end{rmk}
\begin{algorithm}[!ht]
	\label{alg:2}
	\LinesNumbered
	\caption{Self-Adaptive Scheme for DoS Resilient Distributed Consensus Control}
	\ForEach{$(i,j)\in \mathcal{E}$}{
		\ForEach{communication attempt $k$}{
			\uIf{attempt is unsuccessful}{
				apply \eqref{eq:adaptive controller} and Algorithm~\ref{alg:1} to the unsuccessful solution\;
			}
			\ElseIf{attempt is successful}{
				design $\varepsilon_{ij}^{k},R_{ij}^{k}$ using \eqref{eq:self adaptive}\;
				calculate  $\theta_{ij}^{k}$ as \eqref{eq:adaptive controller} and $u_{ij}^{k}=u'_{ij}$ as \eqref{eq:pro2}\;
			}
		}
	}
\end{algorithm}

\begin{figure}[!ht]
\centering
\includegraphics[width=\columnwidth]{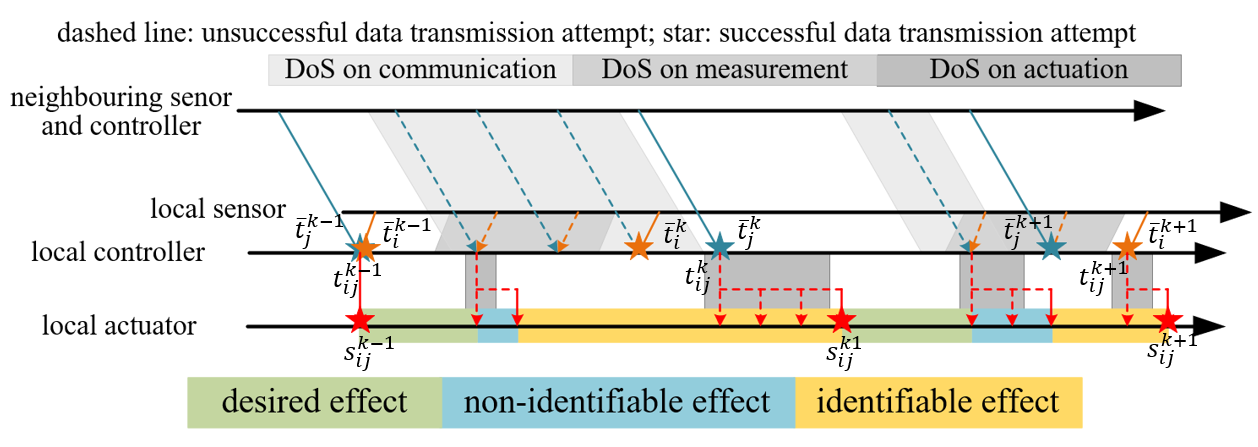}
\caption{Sequential control scenarios under multi-layer DoS attacks.}
\label{fig:timeline}
\end{figure}

\begin{rmk}\label{rmk:conservative_note}
    Under \cref{cor:dynamic edge-based self-triggered strategy}, the adverse effects of multi-layer DoS attacks can be classified as ``identifiable" and ``non-identifiable" depending on the extent to which the convervativeness of global consensus criteria \eqref{eq:stability criteria} can be mitigated, as shown in \cref{fig:timeline}. More specifically, the ``identifiable" means those DoS attacks can be noticed before control command calculation by the definition of \eqref{eq:timestamp} (e.g., communication and measurement DoS), while the ``non-identifiable" means the actuated commands are not updated as desired due to DoS attacks that block the next actuation attempt (e.g., actuation DoS). The ``non-identifiable" effects come always with actuation DoS attacks and are mitigated by using \cref{prop:for self-adaptive}, which brings extra conservativeness. Besides the desired effects, such separation of identifiable and non-identifiable effects can effectively avoid the over conservative design using the fully worst scenario owing to intensive DoS attacks are a low-frequency event.
\end{rmk}

{
\begin{rmk}\label{rmk:comparison}
     Compared to \cite{senejohnny_jamming-resilient_2018,feng_secure_2019,xu2019distributed}, the main contributions of the proposed method are: 1) consideration of  the multi-layer DoS attacks in all channels of local measurement, neighbouring communication and control actuation, 2) consideration of asynchronous data collection and processing, as major significance, to ensure consensus properties in the presence of multi-layer DoS attacks, 3) the proposed adaptive scheme can significantly reduce the conservativeness involved in the algorithm \cite{senejohnny_jamming-resilient_2018}. These contributions lead to a dedicated resilient control design with rigorous analysis for resilience guarantees. To show the superior of the proposed method, comprehensive comparisons with \cite{feng_secure_2019,xu2019distributed,senejohnny_jamming-resilient_2018,de_persis_robust_2013} will be provided in \cref{sec:5.a}.
\end{rmk}
}

\section{Results}\label{sec:5}
\begin{figure}[!ht]
	\centering
	\includegraphics[width=\columnwidth]{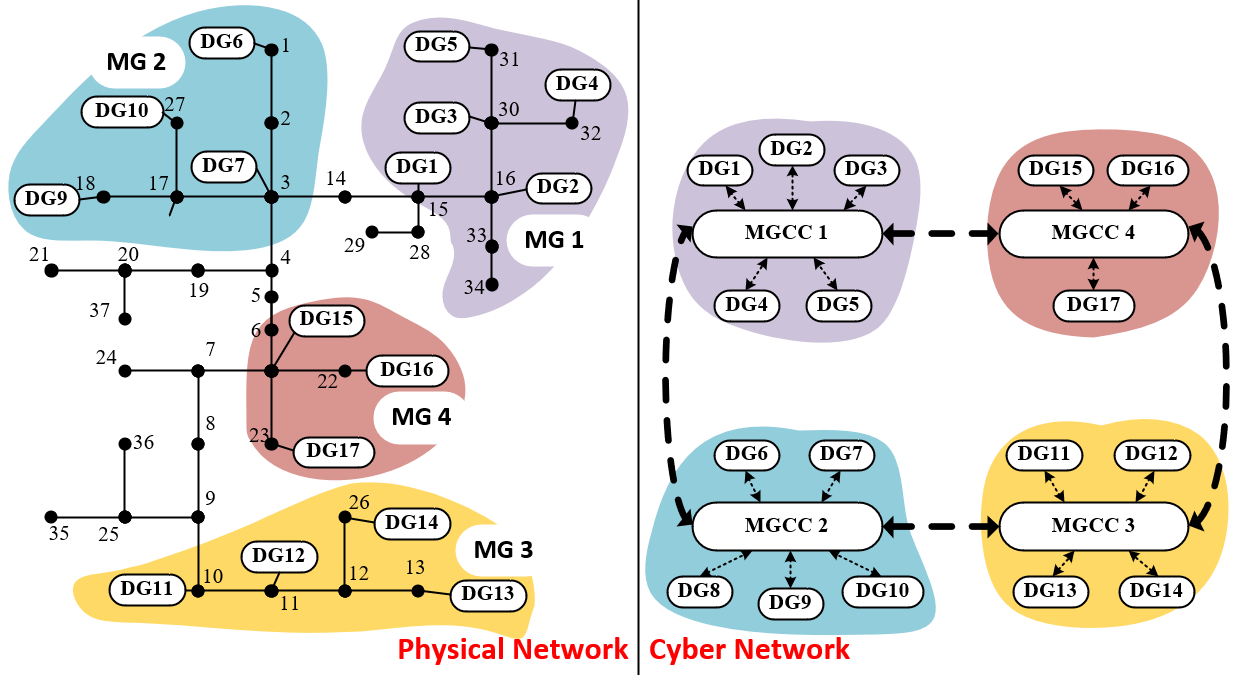}\\[-2ex]
	\caption{A networked MGs topology modified by IEEE 37 bus test system.}
	\label{fig:topo}
\end{figure}
\begin{figure}[!ht]
	\centering
	\includegraphics[width=\columnwidth]{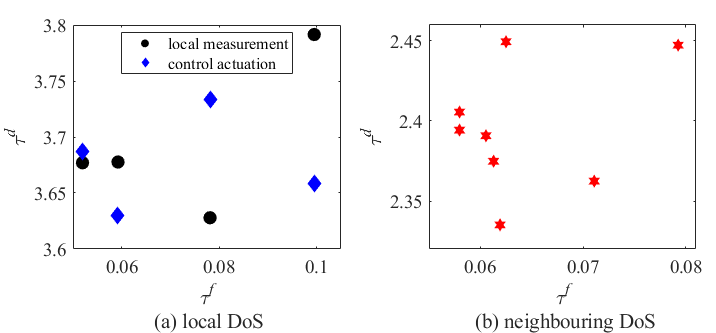}\\[-2ex]
	\caption{$\tau^{f}$ and $\tau^{d}$ values among networked MGs: (a) measurement and control actuation; (b) neighbouring communication.}
	\label{fig:mg_attack}
\end{figure}

\begin{table}[!ht]
	\centering
	\caption{Power Ratings of DGs}
	\label{table:parameter}
	\renewcommand\arraystretch{1.3}
	\resizebox{0.48\textwidth}{!}{%
		\begin{tabular}{l|l|l|l|l|l|l|l}
			\hline
			\multicolumn{2}{c|}{MG 1} & \multicolumn{2}{c|}{MG 2} & \multicolumn{2}{c|}{MG 3} & \multicolumn{2}{c}{MG 4} \\ \hline
			DG 1 & 20 kW & DG 6  & 20 kW & DG 11 & 15 kW & DG 15 & 10 kW \\
			DG 2 & 15 kW & DG 7  & 20 kW & DG 12 & 20 kW & DG 16 & 10 kW \\
			DG 3 & 15 kW & DG 8  & 15 kW & DG 13 & 20 kW & DG 17 & 15 kW \\
			DG 4 & 15 kW & DG 9  & 15 kW & DG 14  & 15 kW &       &       \\
			DG 5 & 15 kW & DG 10 & 10 kW &       &       &       &       \\ \hline
		\end{tabular}%
	}
\end{table}
{To verify the effectiveness of the proposed DoS resilient control of networked MGs, a modified IEEE 37 nodes system \cite{ge_extended-state-observer-based_2020} with four MGs is established in MATLAB/Simulink as shown in \cref{fig:topo}. The network topology follows $\mathcal{A} = \begin{footnotesize}\left[\begin{array}{cccc}
         0 &1 &0 &1  \\
         1 &0 &1 &0  \\
         0 &1 &0 &1  \\
         1 &0 &1 &0
         \end{array}\right]\end{footnotesize}$,
which satisfies the consensus requirement discussed in \cref{sec:2.1} .}
Each MG incorporates several inverter-based DGs, the power ratings of which are detailed in \cref{table:parameter}. In the simulation, the proposed secondary controller is activated at $t=5\,$s, and before only the primary controller is used, which tends to lead to larger frequency synchronous deviations. Furthermore, the load changes (prevalent in the power networks) are introduced at $t=30\,$s and $t=45\,$s, respectively. Finally, multi-layer DoS attacks acting on local and neighbouring links of the power network are illustrated in Fig.~\ref{fig:mg_attack}.

\begin{figure*}[!ht]
	\centering
	\includegraphics[width=0.95\textwidth]{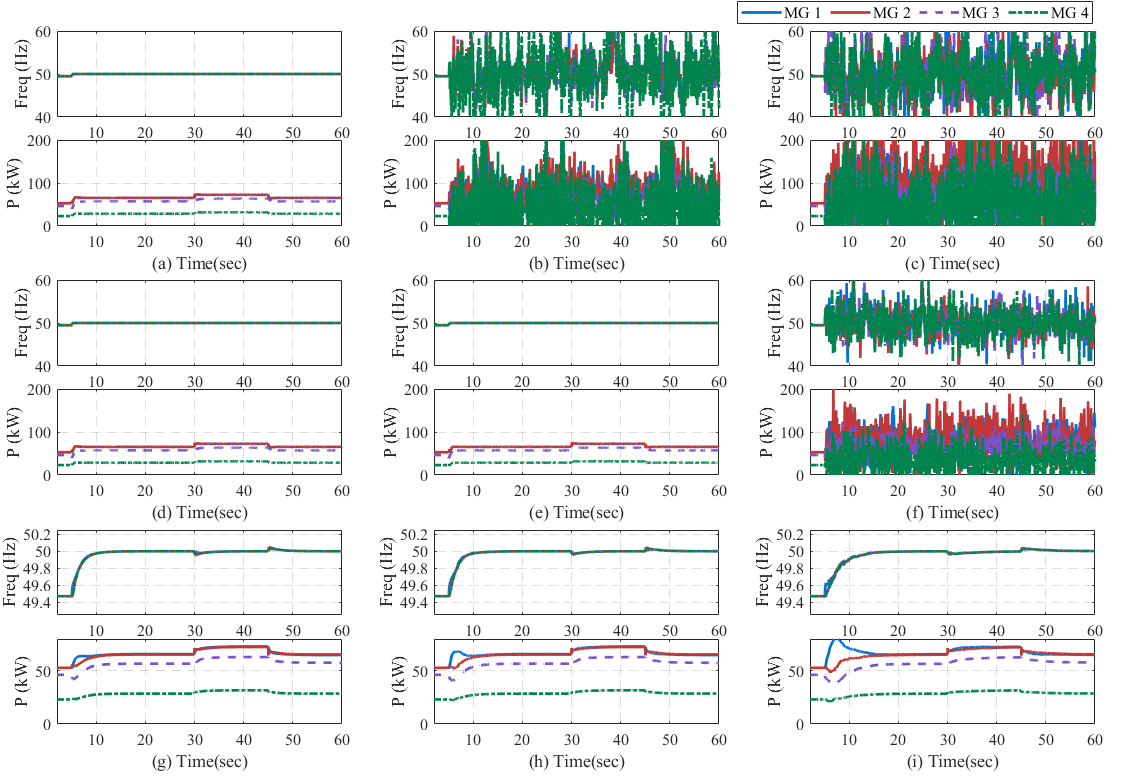}\\[-2ex]
	\caption{{Performance evaluation of frequency synchronisation and active power sharing. 1st row, i.e., (a), (b), (c) are using \eqref{eq:controller} designed without considering any DoS attacks \cite{de_persis_robust_2013}; 2nd row, i.e., (d), (e), (f) are using ternary control \eqref{eq:controller} designed only considering neighbouring DoS attacks \cite{feng_secure_2019,xu2019distributed,senejohnny_jamming-resilient_2018}; 3rd row, i.e., (g), (h), (i) are using the proposed resilient control designed considering multi-layer DoS attacks; 1st column, i.e., (a), (d), (g): none DoS attacks exist; 2nd column, i.e., (b), (e), (h): only communication DoS attacks exist; 3rd column, i.e., (c), (f), (i): multi-layer DoS attacks exist.}}
	\label{fig:9}
\end{figure*}
\begin{figure*}[!htb]
	\centering
	\includegraphics[width=\textwidth]{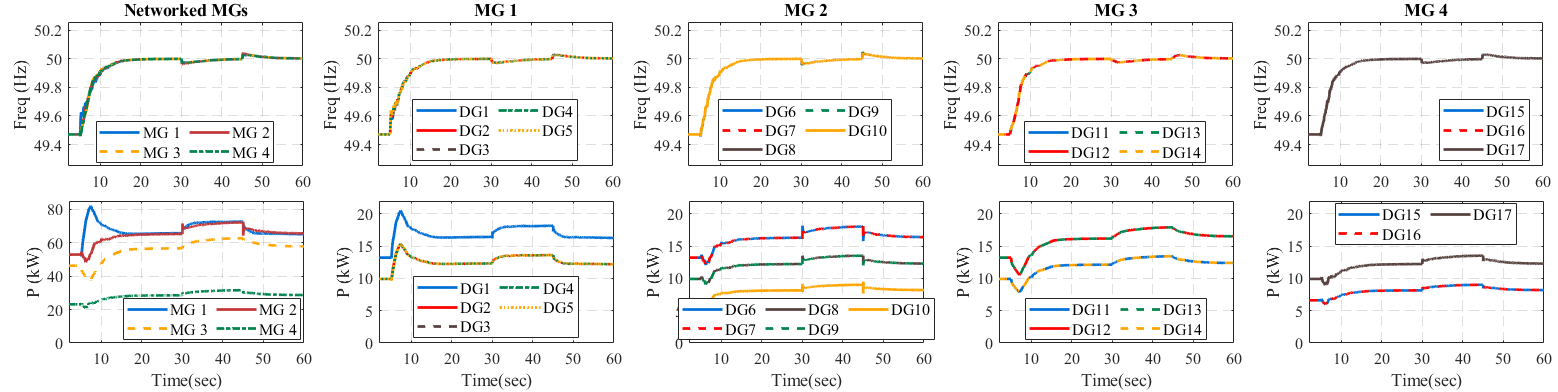}\\[-2ex]
	\caption{Frequency synchronisation and active power sharing inside MGs.}
	\label{fig:A3}
\end{figure*}

\begin{figure*}[!htb]
	\centering
	\includegraphics[width=0.95\textwidth]{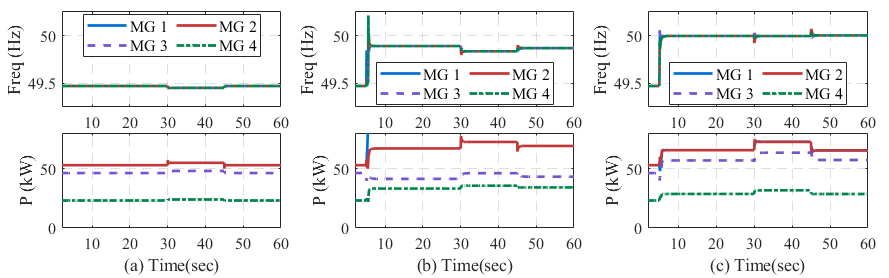}\\[-2ex]
	\caption{Conservativeness validation of \cref{thm:stability criteria}. (a): intensive DoS attacks using controller satisfying \cref{thm:stability criteria}; (b) and (c): less intensive DoS attack using controllers satisfying \cref{thm:stability criteria} and \cref{cor:dynamic edge-based self-triggered strategy}, respectively.}
	\label{fig:3}
\end{figure*}

\begin{figure*}[!htb]
	\centering
	\includegraphics[width=\textwidth]{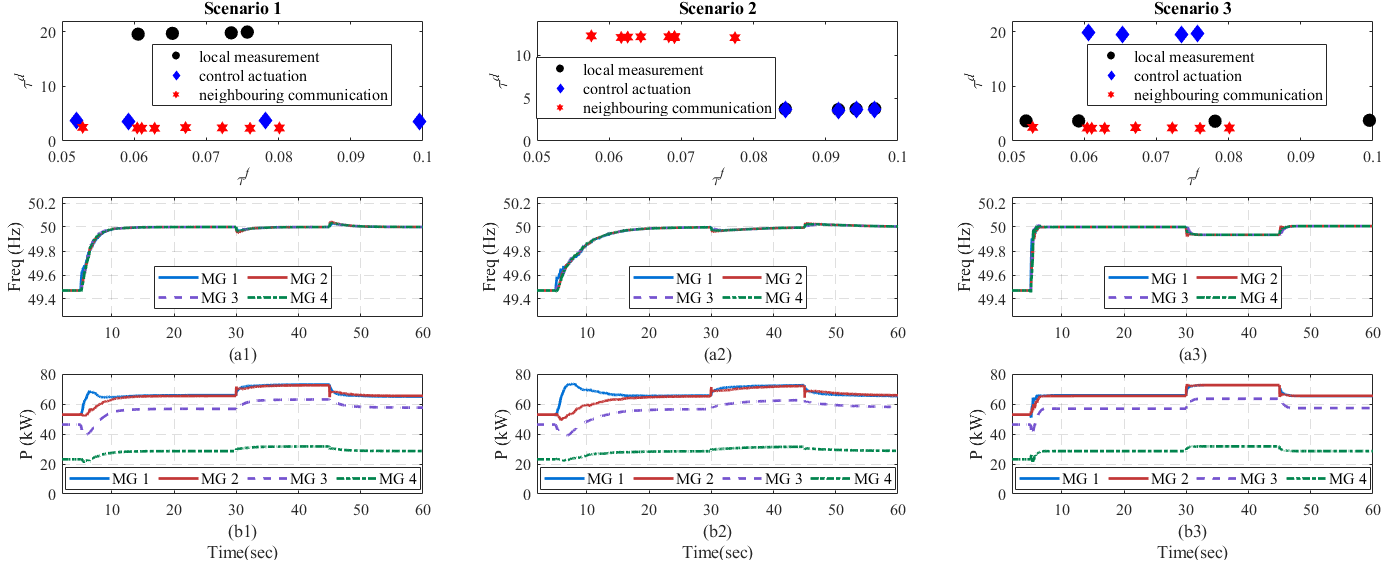}\\[-2ex]
	\caption{Performance comparisons with decreased DoS attacks on three type of channels separately: measurement (Scenario 1, 1st column), communication (Scenario 2, 2nd column) and actuation (Scenario 3, 3rd column).}
	\label{fig:C1}
\end{figure*}

\subsection{Validation of the Proposed Method}\label{sec:5.a}
{To show the impact of multi-layer DoS attacks and the performance of the proposed resilient secondary control strategy which is based on \cref{cor:dynamic edge-based self-triggered strategy}, we compare the performance with existing methods~\cite{feng_secure_2019,xu2019distributed,senejohnny_jamming-resilient_2018,de_persis_robust_2013}. The results are shown in \cref{fig:9}, where each row corresponds to a typical controller and the three columns (from left to right) indicate the three simulation cases of different DoS attacks.}
As it can be seen, control performance deteriorates under either neighbouring DoS attacks or local DoS attacks (see (a) to (b)), and the degradation becomes more significant when local DoS attacks are introduced (see (b) to (c)). Considering only the neighbouring-communication-attack can not nullify the effects of local DoS attacks (see (e) to (f)).
The resulting undesired oscillations may trigger the power grid protection mechanism, and consequently, lead to large-scale load shedding or power outage. Hence, the resilience against multi-layer DoS attacks is of great significance for enhancing the reliability of the networked MGs. The results presented in the third row (i.e., (g), (h) and (i)) show that system resilience is preserved by the proposed DoS-resilient control method although the multi-layer DoS attacks slow down the frequency convergence speed. Moreover, frequency synchronisation and active power sharing are shown by equivalence inside each MG in \cref{fig:A3}, where the accuracy is also guaranteed in a hierarchical framework. Take MG 2 as an example, the active power sharing is kept at all stages by a fixed ratio $4:4:3:3:2$, as specified by their power ratings.

\subsection{Benefits of the Self-Adaptive Scheme}
Under the DoS attacks of \cref{fig:mg_attack}, we evaluate the performance of the controller designed in line with the global consensus criteria (see \cref{thm:stability criteria}), which considers the worst scenario of DoS attacks by PoDF. The results are shown in \cref{fig:3}(a). In contrast to \cref{fig:9}(i) that is obtained using the self-adaptive scheme, the steady state consensus error in \cref{fig:3}(a) is much greater due to the fact that the sensitivity parameter, $\varepsilon$, has to be set to a conservative value $\varepsilon=1.2624 \,(\Phi_{\mathcal{I}=\mathcal{I}0}^{\max}=0.0526)$ to satisfy the global design criterion \cref{eq:stability criteria}. If DoS attacks become less severe and intensive, after re-designing the the sensitivity parameter, the consensus accuracy is improved for both control designs, as can be seen in \cref{fig:3}(b) and \cref{fig:3}(c). However, enhanced consensus accuracy is guaranteed in both cases by the less conservative design criteria given in \cref{cor:dynamic edge-based self-triggered strategy}.

\subsection{Impacts of Attacks in Different Channels}
The proposed DoS-resilient control framework gives different mitigation methods for identifiable and non-identifiable DoS attacks as described in \cref{rmk:conservative_note}. In order to evaluate the impacts of both types of attacks and to what extent each attack can be mitigated, we successively decrease the frequency and duration for measurement, communication or actuation DoS attacks based on the original setting given in \cref{fig:mg_attack}. The resulting multi-layer DoS attacks are characterised in the first row of \cref{fig:C1}. The corresponding performances of each scenario are shown in 2nd and 3rd rows of the same column. As discussed in \cref{rmk:conservative_note}, the mitigation of the non-identifiable attacks is more conservative compare to that of identifiable ones. This is explicitly reflected in \cref{fig:C1}, as the extenuation (by frequency and duration reduction) of the actuation attacks (which certainly bring non-identifiable effects) yields the most noticeable improvements in terms of frequency tracking among the three cases (see Scenario 3). In other words, under the proposed resilient self-triggered method based on \cref{cor:dynamic edge-based self-triggered strategy}, a sequence of DoS attack that acts on actuation channels has the most significant impact on the control performance, therefore, it is more beneficial to harden cyber security of actuation channels compared to the other two.

\section{Conclusion}\label{sec:6}
In this paper, we propose a DoS resilient distributed self-triggered control method of networked MGs systems. Multi-layer DoS attacks on different channels of data flow are considered: DoS attacks on neighbouring communication, measurement and control actuation channels. The quantitative description of such attacks, named by PoDF is employed to analyse the global stability criteria and convergence time of the consensus evolution. Then, the conservativeness induced by control design in the worst case is overcome by a self-adaptive scheme which classifies effects of DoS attacks into identifiable and non-identifiable parts. Through simulations conducted by MATLAB/Simulink, the effectiveness of such a multi-layer-DoS resilient strategy is illustrated with separate analysis of DoS attacks on local or neighbouring data transmissions.

{In this paper, we assume all channels in information systems vulnerable to DoS attacks. However, in some cases, if the attacker has limited resources, there is an optimisation problem to allocate attack resources to maximise/minimise the consequences, which in turn suggests an optimization problem for the defender to allocate the defence resources, which is, however, out of the scope of this paper and  will be discussed in other future works.} In addition, this paper only investigates the system dynamics that are modelled by the first-order, and it is interesting to conduct research on more accurately modelled networked MGs. Moreover, cybersecurity issues do not only include DoS, thus deception attacks such as false data injection (FDI) will be considered in the future.

%
%

\appendices
\section{Proof}

\subsection{Proof of Theorem~\ref{thm:local stability criteria}}\label{app:thm2}
\begin{IEEEproof}
	From the proof of Theorem~\ref{thm:stability criteria}, the inequality \eqref{eq:Dij} and \eqref{eq:Dij_negative} can be replaced by
	\begin{align}
		\left\{\begin{aligned}
			\begin{aligned}
				D_{ij}(t)
				&\geq D_{ij}(\bar{t}_{ij}^{k})(1-\frac{1}{2R_{ij}})-d_{i}(\Phi_{i}+2\Phi_{i0})\\
				&\quad-d_{j}(\Phi_{j}+2\Phi_{i0}),\mbox{ if }D_{ij}(\bar{t}_{ij}^{k})\geq\varepsilon_{ij}
			\end{aligned}\\
			\begin{aligned}
				D_{ij}(t)
				&\leq D_{ij}(\bar{t}_{ij}^{k})(1-\frac{1}{2R_{ij}})+d_{i}(\Phi_{i}+2\Phi_{i0})\\
				&\quad+d_{j}(\Phi_{j}+2\Phi_{i0}),\mbox{ if }D_{ij}(\bar{t}_{ij}^{k})\leq-\varepsilon_{ij}
			\end{aligned}
		\end{aligned}\right.
	\end{align}
	Then, \eqref{eq:stability proof} can be replaced by
	\begin{align}
		\begin{aligned}
			\dot{V}(t)
			&\leq -\frac{1}{2}\sum_{(i,j)\in\mathcal{E}:\mathcal{S}}\bigg[\varepsilon_{ij}(1-\frac{1}{2R_{ij}})\\
			&\quad-d_{i}(\Phi_{i}+2\Phi_{i0})-d_{j}(\Phi_{j}+2\Phi_{i0})\bigg]<0
		\end{aligned}
	\end{align}
	which shows the convergence using \eqref{eq:local stability criteria} in Theorem~\ref{thm:local stability criteria}. Thus, the secure consensus \eqref{eq:convergence set} is achieved.
\end{IEEEproof}

\subsection{Proof of Proposition~\ref{prop:for self-adaptive}}\label{app:prop2}
\begin{IEEEproof}
	By the inequality $s_{ij}^{k+1}-t_{ij}^{k+1}=t_{i0}^{k+1}\leq\Phi_{i0}$ and $t_{ij}^{k+1}-s_{ij}^{k}=\vartheta_{ij}^{k}$, if $\mathrm{sign}_\varepsilon\left(D_{ij}(\bar{t}_{ij}^{k})\right)=1\Rightarrow u'_{ij}(t)>0,t\in[s_{ij}^{k},s_{ij}^{k+1})$,
	\begin{align}
		\int_{s_{ij}^{k}}^{s_{ij}^{k+1}}u'_{ij}(t)dt\leq\int_{s_{ij}^{k}}^{t_{ij}^{k+1}+\Phi_{i0}}u'_{ij}(t)dt
		\label{eq:lemma_proof_1}
	\end{align}
	if $\mathrm{sign}_\varepsilon\left(D_{ij}(\bar{t}_{ij}^{k})\right)=-1\Rightarrow u'_{ij}(t)<0,t\in[s_{ij}^{k},s_{ij}^{k+1})$,
	\begin{align}
		\int_{s_{ij}^{k}}^{s_{ij}^{k+1}}u'_{ij}(t)dt\geq\int_{s_{ij}^{k}}^{t_{ij}^{k+1}+\Phi_{i0}}u'_{ij}(t)dt
		\label{eq:lemma_proof_2}
	\end{align}
	Combining \eqref{eq:lemma_proof_1} and \eqref{eq:lemma_proof_2}, the contribution of control actuation during $[s_{ij}^{k},s_{ij}^{k+1})$ is limited:
	\begin{align}
		\begin{aligned}
			&\int_{s_{ij}^{k}}^{s_{ij}^{k+1}}\left|u'_{ij}(t)\right|dt\leq\left|\mathrm{sign}_\varepsilon\left(D_{ij}(\bar{t}_{ij}^{k})\right)\right|\vartheta_{ij}^{k}\\
			&\qquad=\int_{s_{ij}^{k}}^{t_{ij}^{k+1}}\left|\mathrm{sign}_\varepsilon\left(D_{ij}(\bar{t}_{ij}^{k})\right)\right|dt + \int_{t_{ij}^{k+1}}^{s_{ij}^{k+1}} 0 \enspace dt
		\end{aligned}\label{eq:lemma_proof_3}
	\end{align}
	Thus, from \eqref{eq:lemma_proof_3}, we can know if $u'_{ij}$ is actuated, it has the equivalent contribution of
	\begin{align*}
		u_{ij}(t)=\left\{\begin{aligned}
			&\mathrm{sign}_\varepsilon\left(D_{ij}(\bar{t}_{ij}^{k})\right),&&s_{ij}^{k}<t<t_{ij}^{k*}\\
			&0,&&t_{ij}^{k*}<t<s_{ij}^{k+1}
		\end{aligned}
		\right.
	\end{align*}
	where $s_{ij}^{k}+\frac{(\vartheta_{ij}^{k})^{2}}{\vartheta_{ij}^{k}+\Phi_{i0}}\leq t_{ij}^{k*}\leq t_{ij}^{k+1}$. In particular, $t_{ij}^{k*}=s_{ij}^{k}+\frac{(\vartheta_{ij}^{k})^{2}}{\vartheta_{ij}^{k}+\Phi_{i0}}$ implies $t_{ij}^{k+1}=s_{ij}^{k+1}$.
\end{IEEEproof}

\subsection{Proof of Corollary \ref{cor:dynamic edge-based self-triggered strategy}}\label{app:cor1}
\begin{IEEEproof}
	If $D_{ij}(\bar{t}_{ij}^{k})\geq \varepsilon_{ij}^{k}$, \eqref{eq:local slack} in Theorem~\ref{thm:stability criteria} can be modified as the following
	\begin{align*}
		\begin{aligned}
			& D_{ij}(t)
			\geq D_{ij}(t_{ij}^{k})-(d_i+d_j)(t-t_{ij}^{k})\\
			&\quad \overset{(c)}{\geq} D_{ij}(\bar{t}_{ij}^{k})\!-\!d_{i}t_{i,i}^{k}\!-\!d_{j}t_{i,j}^{k}\!-\!(d_i\!+\!d_j)(t_{i0}^{k}\!+\!\vartheta_{ij}^{k})\!-\!0\!\times\!\Phi_{i0}\\
			&\quad = D_{ij}(\bar{t}_{ij}^{k})(1-\frac{1}{2R_{ij}^{k}})-d_{i}(t_{i,i}^{k}+t_{i0}^{k})-d_{j}(t_{i,j}^{k}+t_{i0}^{k})
		\end{aligned}
	\end{align*}
	where $(c)$ comes from Proposition~\ref{prop:for self-adaptive}. Followed by the similar process as \eqref{eq:Dij}-\eqref{eq:stability proof}, we obtain $\dot{V}(t)<0$ remains guaranteed with \eqref{eq:self adaptive}. Similarly, secure consensus \eqref{eq:convergence set} is achieved.
\end{IEEEproof}


\bibliographystyle{IEEEtran}
\bibliography{references}

\begin{thebibliography}{10}
\providecommand{\url}[1]{#1}
\csname url@samestyle\endcsname
\providecommand{\newblock}{\relax}
\providecommand{\bibinfo}[2]{#2}
\providecommand{\BIBentrySTDinterwordspacing}{\spaceskip=0pt\relax}
\providecommand{\BIBentryALTinterwordstretchfactor}{4}
\providecommand{\BIBentryALTinterwordspacing}{\spaceskip=\fontdimen2\font plus
\BIBentryALTinterwordstretchfactor\fontdimen3\font minus
  \fontdimen4\font\relax}
\providecommand{\BIBforeignlanguage}[2]{{%
\expandafter\ifx\csname l@#1\endcsname\relax
\typeout{** WARNING: IEEEtran.bst: No hyphenation pattern has been}%
\typeout{** loaded for the language `#1'. Using the pattern for}%
\typeout{** the default language instead.}%
\else
\language=\csname l@#1\endcsname
\fi
#2}}
\providecommand{\BIBdecl}{\relax}
\BIBdecl

\bibitem{creutzig2014catching}
F.~Creutzig, J.~C. Goldschmidt, P.~Lehmann, E.~Schmid, F.~von Bl{\"u}cher,
  C.~Breyer, B.~Fernandez, M.~Jakob, B.~Knopf, S.~Lohrey \emph{et~al.},
  ``Catching two european birds with one renewable stone: Mitigating climate
  change and eurozone crisis by an energy transition,'' \emph{Renewable and
  Sustainable Energy Reviews}, vol.~38, pp. 1015--1028, 2014.

\bibitem{wang2022electrifying}
J.~Wang, R.~El~Kontar, X.~Jin, and J.~King, ``Electrifying high-efficiency
  future communities: Impact on energy, emissions, and grid,'' \emph{Advances
  in Applied Energy}, vol.~6, p. 100095, 2022.

\bibitem{wang2020sustainable}
J.~Wang and X.~Lu, ``Sustainable and resilient distribution systems with
  networked microgrids,'' \emph{Proceedings of the IEEE}, vol. 108, no.~2, pp.
  238--241, 2020.

\bibitem{gholami2021stability}
A.~Gholami and A.~Sun, ``Stability of multi-microgrids: New certificates,
  distributed control, and braess's paradox,'' \emph{IEEE Transactions on
  Control of Network Systems}, 2021.

\bibitem{ge2022resilience}
P.~Ge, F.~Teng, C.~Konstantinou, and S.~Hu, ``A resilience-oriented
  centralised-to-decentralised framework for networked microgrids management,''
  \emph{Applied Energy}, vol. 308, p. 118234, 2022.

\bibitem{pasqualetti_control-theoretic_2015}
F.~Pasqualetti, F.~Dorfler, and F.~Bullo, ``Control-{Theoretic} {Methods} for
  {Cyberphysical} {Security}: {Geometric} {Principles} for {Optimal}
  {Cross}-{Layer} {Resilient} {Control} {Systems},'' \emph{IEEE Control Systems
  Magazine}, vol.~35, no.~1, pp. 110--127, Feb. 2015.

\bibitem{ge_eventtriggered_2020}
P.~Ge, B.~Chen, and F.~Teng, ``Event-triggered distributed model predictive
  control for resilient voltage control of an islanded microgrid,''
  \emph{International Journal of Robust and Nonlinear Control}, vol.~31, no.~6,
  pp. 1979--2000, 2021.

\bibitem{ge_resilient_2021}
P.~Ge, Y.~Zhu, T.~C. Green, and F.~Teng, ``\BIBforeignlanguage{en}{Resilient
  {Secondary} {Voltage} {Control} of {Islanded} {Microgrids}: {An}
  {ESKBF}-{Based} {Distributed} {Fast} {Terminal} {Sliding} {Mode} {Control}
  {Approach}},'' \emph{\BIBforeignlanguage{en}{IEEE Transactions on Power
  Systems}}, vol.~36, no.~2, pp. 1059--1070, Mar. 2021.

\bibitem{deng2021distributed}
C.~Deng, F.~Guo, C.~Wen, D.~Yue, and Y.~Wang, ``Distributed resilient secondary
  control for dc microgrids against heterogeneous communication delays and dos
  attacks,'' \emph{IEEE Transactions on Industrial Electronics}, 2021.

\bibitem{lian2021distributed}
Z.~Lian, F.~Guo, C.~Wen, C.~Deng, and P.~Lin, ``Distributed resilient optimal
  current sharing control for an islanded dc microgrid under dos attacks,''
  \emph{IEEE Transactions on Smart Grid}, vol.~12, no.~5, pp. 4494--4505, 2021.

\bibitem{danzi2019software}
P.~Danzi, M.~Angjelichinoski, {\v{C}}.~Stefanovi{\'c}, T.~Dragi{\v{c}}evi{\'c},
  and P.~Popovski, ``Software-defined microgrid control for resilience against
  denial-of-service attacks,'' \emph{IEEE Transactions on Smart Grid}, vol.~10,
  no.~5, pp. 5258--5268, 2019.

\bibitem{liu2019stochastic}
S.~Liu, Z.~Hu, X.~Wang, and L.~Wu, ``Stochastic stability analysis and control
  of secondary frequency regulation for islanded microgrids under random denial
  of service attacks,'' \emph{IEEE Transactions on Industrial Informatics},
  vol.~15, no.~7, pp. 4066--4075, 2019.

\bibitem{hu2022resilient}
S.~Hu, F.~Yang, S.~Gorbachev, D.~Yue, V.~Kuzin, and C.~Deng, ``Resilient
  control design for networked dc microgrids under time-constrained dos
  attacks,'' \emph{ISA transactions}, 2022.

\bibitem{wan2021distributed}
Y.~Wan, C.~Long, R.~Deng, G.~Wen, X.~Yu, and T.~Huang, ``Distributed
  event-based control for thermostatically controlled loads under hybrid cyber
  attacks,'' \emph{IEEE Transactions on Cybernetics}, vol.~51, no.~11, pp.
  5314--5327, 2021.

\bibitem{zhang2021attack}
B.~Zhang, C.~Dou, D.~Yue, J.~H. Park, and Z.~Zhang, ``Attack-defense
  evolutionary game strategy for uploading channel in consensus-based secondary
  control of islanded microgrid considering dos attack,'' \emph{IEEE
  Transactions on Circuits and Systems I: Regular Papers}, vol.~69, no.~2, pp.
  821--834, 2022.

\bibitem{hu2020resilient}
Z.~Hu, S.~Liu, W.~Luo, and L.~Wu, ``Resilient distributed fuzzy load frequency
  regulation for power systems under cross-layer random denial-of-service
  attacks,'' \emph{IEEE Transactions on Cybernetics}, 2020.

\bibitem{hu2020attack}
S.~Hu, P.~Yuan, D.~Yue, C.~Dou, Z.~Cheng, and Y.~Zhang, ``Attack-resilient
  event-triggered controller design of dc microgrids under dos attacks,''
  \emph{IEEE Transactions on Circuits and Systems I: Regular Papers}, vol.~67,
  no.~2, pp. 699--710, 2020.

\bibitem{liu2021resilient}
X.-K. Liu, C.~Wen, Q.~Xu, and Y.-W. Wang, ``Resilient control and analysis for
  dc microgrid system under dos and impulsive fdi attacks,'' \emph{IEEE
  Transactions on Smart Grid}, vol.~12, no.~5, pp. 3742--3754, 2021.

\bibitem{chlela2018fallback}
M.~Chlela, D.~Mascarella, G.~Joos, and M.~Kassouf, ``Fallback control for
  isochronous energy storage systems in autonomous microgrids under
  denial-of-service cyber-attacks,'' \emph{IEEE transactions on smart grid},
  vol.~9, no.~5, pp. 4702--4711, 2018.

\bibitem{chen2022multi}
P.~Chen, S.~Liu, B.~Chen, and L.~Yu, ``Multi-agent reinforcement learning for
  decentralized resilient secondary control of energy storage systems against
  dos attacks,'' \emph{IEEE Transactions on Smart Grid}, 2022.

\bibitem{heemels_introduction_2012}
W.~Heemels, K.~Johansson, and P.~Tabuada, ``\BIBforeignlanguage{en}{An
  introduction to event-triggered and self-triggered control},'' in
  \emph{\BIBforeignlanguage{en}{2012 {IEEE} 51st {IEEE} {Conference} on
  {Decision} and {Control} ({CDC})}}.\hskip 1em plus 0.5em minus 0.4em\relax
  Maui, HI, USA: IEEE, Dec. 2012, pp. 3270--3285.

\bibitem{dimarogonas_distributed_2012}
D.~V. Dimarogonas, E.~Frazzoli, and K.~H. Johansson, ``Distributed
  event-triggered control for multi-agent systems,'' \emph{IEEE Transactions on
  Automatic Control}, vol.~57, no.~5, pp. 1291--1297, 2011.

\bibitem{feng_secure_2019}
Z.~Feng and G.~Hu, ``Secure cooperative event-triggered control of linear
  multiagent systems under dos attacks,'' \emph{IEEE Transactions on Control
  Systems Technology}, vol.~28, no.~3, pp. 741--752, 2020.

\bibitem{xu2019distributed}
W.~Xu, G.~Hu, D.~W. Ho, and Z.~Feng, ``Distributed secure cooperative control
  under denial-of-service attacks from multiple adversaries,'' \emph{IEEE
  Transactions on Cybernetics}, vol.~50, no.~8, pp. 3458--3467, 2019.

\bibitem{senejohnny_jamming-resilient_2018}
D.~Senejohnny, P.~Tesi, and C.~D. Persis, ``\BIBforeignlanguage{en}{A
  {Jamming}-{Resilient} {Algorithm} for {Self}-{Triggered} {Network}
  {Coordination}},'' \emph{\BIBforeignlanguage{en}{IEEE Transactions on Control
  of Network Systems}}, vol.~5, no.~3, p.~10, 2018.

\bibitem{de_persis_robust_2013}
C.~De~Persis and P.~Frasca, ``Robust {Self}-{Triggered} {Coordination} {With}
  {Ternary} {Controllers},'' \emph{IEEE Transactions on Automatic Control},
  vol.~58, no.~12, pp. 3024--3038, Dec. 2013.

\bibitem{backhaus2016networked}
S.~N. Backhaus, L.~Dobriansky, S.~Glover, C.-C. Liu, P.~Looney, S.~Mashayekh,
  A.~Pratt, K.~Schneider, M.~Stadler, M.~Starke \emph{et~al.}, ``Networked
  microgrids scoping study,'' Los Alamos National Lab.(LANL), Los Alamos, NM
  (United States), Tech. Rep., 2016.

\bibitem{chen2021aggregated}
J.~Chen, M.~Liu, and F.~Milano, ``Aggregated model of virtual power plants for
  transient frequency and voltage stability analysis,'' \emph{IEEE Transactions
  on Power Systems}, vol.~36, no.~5, pp. 4366--4375, 2021.

\bibitem{roos2020aggregation}
M.~H. Roos, P.~H. Nguyen, J.~Morren, and J.~Slootweg, ``Aggregation of
  component-based grid-feeding der and load models for simulation of microgrid
  islanding transients,'' \emph{Electric Power Systems Research}, vol. 189, p.
  106759, 2020.

\bibitem{dorfler2013kron}
F.~Dorfler and F.~Bullo, ``Kron reduction of graphs with applications to
  electrical networks,'' \emph{IEEE Transactions on Circuits and Systems I:
  Regular Papers}, vol.~60, no.~1, pp. 150--163, 2013.

\bibitem{bidram_multiobjective_2014}
A.~Bidram, A.~Davoudi, and F.~L. Lewis, ``A multiobjective distributed control
  framework for islanded ac microgrids,'' \emph{IEEE Transactions on industrial
  informatics}, vol.~10, no.~3, pp. 1785--1798, 2014.

\bibitem{dehkordi2017distributed}
N.~M. Dehkordi, N.~Sadati, and M.~Hamzeh, ``Distributed robust finite-time
  secondary voltage and frequency control of islanded microgrids,'' \emph{IEEE
  Transactions on Power Systems}, vol.~32, no.~5, pp. 3648--3659, 2017.

\bibitem{ren2005survey}
W.~Ren, R.~W. Beard, and E.~M. Atkins, ``A survey of consensus problems in
  multi-agent coordination,'' in \emph{Proceedings of the 2005, American
  Control Conference, 2005.}\hskip 1em plus 0.5em minus 0.4em\relax IEEE, 2005,
  pp. 1859--1864.

\bibitem{ge_extended-state-observer-based_2020}
P.~Ge, X.~Dou, X.~Quan, Q.~Hu, W.~Sheng, Z.~Wu, and W.~Gu,
  ``Extended-{State}-{Observer}-{Based} {Distributed} {Robust} {Secondary}
  {Voltage} and {Frequency} {Control} for an {Autonomous} {Microgrid},''
  \emph{IEEE Transactions on Sustainable Energy}, vol.~11, no.~1, pp. 195--205,
  Jan. 2020.

\end{thebibliography}

\end{document}